\documentclass[a4paper,fleqn,usenatbib]{mnras}
\usepackage{amsmath}
\usepackage{xcolor}
\usepackage{amssymb}
\usepackage{graphicx}
\usepackage{ulem}
\usepackage{cancel}
\usepackage{hyperref}
\usepackage{color}
\usepackage{txfonts}
\usepackage{ae,aecompl}

\usepackage{multirow}

\title[Global stability of self-gravitating discs in modified gravity (MOG)]{Global stability of self-gravitating discs in modified gravity\thanks{By modified gravity we mean a scalar-tensor-vector gravity known as MOG in the literature.}}
\author[N. Ghafourian, M. Roshan]{
Neda Ghafourian,
Mahmood Roshan\thanks{E-mail: mroshan@um.ac.ir}
\\
Department of Physics, Ferdowsi University of Mashhad, P.O. Box 1436, Mashhad, Iran\\
}
\date{Accepted XXX. Received YYY; in original form ZZZ}

\pubyear{2017}

\begin{document}
\label{firstpage}
\pagerange{\pageref{firstpage}--\pageref{lastpage}}
\maketitle
\begin{abstract}
We study the global stability of a self-gravitating disc in the context of Modified Gravity (MOG) using N-body simulations. This theory is a relativistic scalar-tensor-vector theory of 
gravity and presented to address the dark matter problem. In the weak field limit MOG possesses two free parameters $\alpha$ and $\mu_0$ which have been already determined using rotation
curve data of spiral galaxies. The evolution of a stellar self-gravitating disc and more specifically the bar instability in MOG is investigated and compared to a 
Newtonian case. Our models have exponential and Mestel-like surface densities as $\Sigma\propto \exp(-r/h)$ and $\Sigma\propto 1/r$. It is found out that, surprisingly, the discs are more stable against the bar mode 
in MOG than in Newtonian gravity. In other words, the bar growth rate is effectively slower than the Newtonian discs. Also we show that both free parameters, i.e. $\alpha$ and $\mu_0$, have stabilising effects. In other words, increase 
in these parameters will decrease the bar growth rate.
\end{abstract}

\section{\small{Introduction}}
\label{introduction}
Early numerical studies showed that if any flat bulge-less self gravitating disc of particles is set under equilibrium of centrifugal force and the Newtonian gravitational force, 
the particles can not maintain their circular motion and in a timescale very short compared with the lifetime of the spiral galaxies, the disc is heated up and its overall configuration changes from a flat disc to  a stellar
bar, for example see \citet{miller}; \citet{ho}. This means that the disc is globally unstable. This fact is known as the bar instability in the literature. 

To tackle this problem \citet{op} in their pioneering paper showed that by adding a spherical rigid component to the system, the evolution of the initial disc to 
the final bar could be controlled. In fact for the first time in the literature they introduced the dark matter halo concept. Historically, the global stability of the disc galaxies played
a key role among the astrophysicists to accept that the dark matter problem, which had already alarmed by \citet{z}, is a serous one and it is not due to the lacks in 
observational equipment. For a comprehensive review of the subject we refer the reader to \citet{sand2010}. After Ostriker \& Peebles pioneer work, several analytical 
and numerical studies have been done to investigate the role of the dark matter halo in the evolution of spiral galaxies, for example see \citet{se81}; \citet{ef};
\citet{at1986}; \citet{at2002} and \citet{se14} for a review of the subject.

It is well understood that dark matter halo stabilises the disc against global perturbations and slows down the growth rate of the stellar bar. 
On the other hand, it should be noted that about 65\% of luminous spiral galaxies are barred \citep{sh}. The presence of bars in the real galactic discs is much
larger than traditionally thought. This fraction is a function of the cosmic redshift $z$, and is smaller at larger redshifts. For example it drops to 20\% at $z=0.8$
\citep{sh}. This means 
that stellar bars have been effectively formed in the spiral galaxies during the past $7$ Gyr. Furthermore, bars exhibit interesting features and play important roles
in the 
evolution of the disc galaxies. They seem to be linked to some engrossing phenomena like secular evolution and pseudo-bulge growth \citep{KK} and galactic 
rings \citep{com}. Furthermore, bars are key ingredients that help to redistribute angular momentum between different components of disc galaxies \citep{at2002}. 
They are also thought to excite spiral arms \citep{toom69,san1976}. They transport gas to the centre of disc galaxies and may trigger AGN activity. Therefore it is 
necessary to emphasise that the dark matter halo is needed to slow down the bar growth and not to totally suppress it.

It is clear that fate and dynamics of stellar bars closely depend on the properties of the dark matter halos. In other words, the bar growth is directly related to the dark matter problem. More specifically, dark matter halos not only moderate the bar instability but also are necessary to explain the flat rotation curves of the spiral galaxies. However dark matter particles have not 
yet been discovered while there are several laboratories which look for these exotic particles using direct and indirect techniques, see \citet{berton}. This fact keeps open
another approach to the problem: i.e. modified gravity. These theories are widely used to address the dark energy (for example see \citealt{cap} for a review
of dark energy models) and the dark matter problem (for example see \citealt{milgrom} and \citealt{fa} for modified Newtonian dynamics (MOND) , \citealt{m2006} for Modified
Gravity (MOG) and \citealt{milli} for a symmetron-like non-minimally coupled scalar-tensor theory ).

Among the modified theories presented to resolve the dark matter problem, MOND (and its corresponding relativistic theory TeVeS; \citealt{bek}) is one of the most successful theories.
The global stability of spiral galaxies in MOND has been investigated in \citet{chris}; \citet{brada} and \citet{ti} using N-body simulations. 
It is found in \citet{chris} and \citet{brada} that disc galaxies are more stable in MOND than in Newtonian gravity. \citet{ti} showed that although 
the bar instability occurs sooner in MOND than in Newtonian gravity, stellar bar weaken in MOND while in the dark matter halo model they continue to grow. 

In this paper, we study the bar instability in MOG using N-body simulations. One may naturally expect that the stability of the spiral galaxies should be explained in 
this theory without any need to 
dark matter halos. In the weak field limit, MOG is able to fit a large number of galaxy rotation curve data 
\citet{br2006}, as well as the X-ray galaxy clusters data \citet{br2006b} without any need of non baryonic matter. Recently it has been claimed in 
\citet{im} that MOG has the
potential to explain merging cluster dynamics, such as Bullet Cluster and Train Wreck
Cluster, without dark matter.  For some cosmological consequences of this theory, we refer the reader to \citet{m2009}; \citet{m2015}; \citet{roepjc} and \citet{jamali}. Furthermore, the local 
stability of disc galaxies in MOG has been already investigated in \citet{ro2015} and the generalised Toomre criterion has been derived. Also for some N-body and numerical studies on the dynamics of disc galaxies in MOG see \citet{aran1} and references therein.

The outline of this paper is as follows: In section \ref{wfl} we briefly discuss the weak filed limit of MOG. In section \ref{gi} we set up the initial conditions and 
describe the
simulation. Results for the exponential model $\Sigma\propto e^{-r/h}$ has been presented in section \ref{res}. We have done some tests on the reliability of the results in the section \ref{T}. In this section we have also presented the results of the stability analysis of the Mestel-like disc. Conclusions are drawn in section \ref{conc}.

\section{\small{Weak field limit of MOG}}
\label{wfl}
MOG is a fully relativistic and covariant generalisation of General Relativity (GR). Despite GR which is a tensor gravity, MOG is a scalar-tensor-vector theory of gravity in which
additional to the metric tensor there are two scalar fields, $\mu(x^\beta)
$ and $ G(x^\beta)$, and also a massive Proca vector field $\phi^\beta$. These extra degrees of freedom enable MOG to explain some astrophysical data related to the dark matter 
problem. In order to study the dynamics of a disc galaxy, we need the weak field limit of this theory. Therefore let us briefly review MOG's modified Poisson equations.
Perturbing the above mentioned fields around their background values and substituting them into the field equations, one can find the modified Poisson equations, for details see 
\citet{m2013}; \citet{mrah2014}; \citet{ro2014}. The equations of motion of spinning and non-spinning test particles in MOG has been investigated in \citet{ro2013}.

In the weak field limit, the test particle's equation of motion is 
\begin{equation}
\frac{d^2 r}{dt^2}=-\nabla \Phi 
\end{equation}
where $\Phi$ is an effective gravitational potential defined as
\begin{equation}
\Phi=\Psi+\chi \phi^0
\end{equation}
in which $\phi^0$ is the zeroth component of the vector field, $\chi$ is a coupling constant, and $\Psi$ and $\phi^0$ satisfy the following differential equations
\begin{equation}
 \nabla^2 \Psi= 4\pi(1+\alpha)G\rho
 \label{p1}
\end{equation}
\begin{equation}
 (\nabla^2-\mu_0^2)\chi\phi^0=-4\pi\alpha G \rho 
\label{p2}
 \end{equation}
where $\chi$ and $\alpha$ are related as $\alpha=\frac{\chi^2}{\omega_0 G}$. Also $\mu_0$ is the background value of the scalar fields $\mu$ and appears as a free parameter, 
$\omega_0$ is a coupling constant, $G$ is the gravitational constant, and $\rho$ is the matter density. It is necessary to mention that MOG possesses two coupling constants 
$\omega_0$ and $\chi$ in its generic action. However in the weak field limit these coupling constants join and form a single free parameter $\alpha$. Observational values of the
free parameters $\alpha$ and $\mu_0$ obtained from rotation curve data of spiral galaxies are $8.89 \pm 0.34$ and $0.042 \pm 0.004 ~\text{kpc}^{-1}$ respectively 
\citet{m2013}. Using equations (\ref{p1}) and (\ref{p2}), the generalised Poisson equation would be
\begin{equation}
 \nabla^2 \Phi(r)=4\pi G \rho +\alpha \mu_0^2 G \int \frac{e^{-\mu_0 |r-r'|}}{|r-r'|} \rho(r')d^3x'
 \label{n1}
\end{equation}
We can simply say that our purpose in this paper is to study the effects of the last term in the right hand side of equation (\ref{n1}) on the global stability of disc galaxies. 
Using equation (\ref{n1}) for a massive test particle, the acceleration law and the gravitational potential can be written as
\begin{equation}
 \ddot{r}=-\frac{G M}{r^2}[1+\alpha-\alpha(1+\mu_0 r)e^{-\mu_0 r}]
 \label{accmog}
\end{equation}
\begin{equation}
 \Phi(r)=-\frac{G M}{r}[1+\alpha-\alpha e^{-\mu_0 r}].
\label{potmog}
 \end{equation}
it is clear that by setting to zero $\alpha$ or $\mu_0$, one can recover the Newtonian gravity. Also since the free parameters are positive, it is easy to show that MOG leads to 
stronger force than the Newtonian case. In fact this is a necessary feature for theories which try to address the dark matter problem.

\section{\small{Global stability of an exponential disc in MOG}}
\label{gi}
In this section we create an initial axi-symmetric disc model with surface density proportional to $exp(-R/h)$ in Newtonian gravity as well as in MOG. Then we explore
their dynamics using an N-body code and concentrate on the bar
instability and the growth rate of the stellar bar. For simplicity, our disc model in Newtonian gravity does not possess a dark matter halo; we will call it a "bare" disc model hereafter. This 
fact will help to compare these theories more clearly. In this sense, this paper follows a same approach as in \citet{op}, where a bare disc model has been compared with a model in
which the gravitational force is modified because of a rigid spherical component. Of course it will be also instructive to compare MOG disc with a more realistic disc model 
including the dark matter halo \citet{gh}.

Let us first introduce two main parameters which will help us to quantify the bar instability in the above mentioned disc models. 
\subsection{Stability parameter $\beta$}
In galactic simulations there is a common parameter to study the stability of the system. For a virialized system we have $T=\frac{1}{2}|W|$, where $T$ is the total kinetic energy 
and $W$ is the gravitational potential energy of the system. The total kinetic 
energy can be decomposed as $T=T_{\text{rot}}+T_{\text{ran}}$. Where $T_{\text{rot}}$ is the total rotational kinetic energy of the particles and $T_{\text{ran}}$ is the total 
kinetic energy associated to the random motions of the particles. 

It is common to define a stability parameter as $ t= \frac{T_{\text{rot}}}{|W|}>0$. Using the Virial theorem one may show that $t$ is always smaller than $\frac{1}{2}$. If this 
parameter decreases with time then the random motions of the system will increase. This can be considered as the outset of the instability. In this paper we use another parameter 
constructed from $t$ as 
\begin{equation}
 \beta=\frac{1}{2t}-1.
\end{equation}
In this case it is easy to show that $\beta=\frac{T_{\text{ran}}}{T_{\text{rot}}}$. It is apparent that a decrease in the ratio of rotational motions to random motions, make $t$ 
parameter to shrink and $\beta$ to grow. Since a bar is a pressure dominated configuration, its formation will increase the $\beta$ parameter. Therefore, this parameter will help
to 
control and interpret the bar instability. Albeit one need to be careful when using this parameter. In fact a growing $\beta$ does not necessarily mean that a bar mode, $m=2$, is 
growing. In principle this parameter can also grow even when there is no excited non-axisymmetric perturbation, $m\neq 0$, in the system. For example when a spherical bulge ($m=0$)
forms in a 
simulation, $\beta$ can in principle increase with time. In other words, although $\beta$ provides a practical tool for exploring the instability, it does not distinguish between 
different modes of perturbations.

Therefore let us introduce other stability parameters which measure the growth rate of a given mode $m$. In this case describing the instability will be more reliable and straightforward.

\subsection{Bar strength and Fourier decomposition}
Various methods have been used to quantify the bar strength. For example \citet{at2003} states that, in a relatively closed system, the ability of the bar to transfer
angular momentum to other galactic components determines the strength of the bar. According to \citet{bc}, in a relatively open system, bar growth and destruction is
linked to the external gas accretion and tidal interactions. Since our disc is an isolated stellar disc, we will study the rate of angular momentum transfer from 
inner part of the disc to the outer parts in order to quantify the bar strength. 

Another way to study the bar strength, is the Fourier intensity amplitude which actually does not specify how influential the bar is on its environment, but 
determines the 
brightness of the bar in contrast to its background. However it could be considered as a measure of bar strength \citealt{cs} and \citealt{ee}. In this method  the various 
components of
the Fourier expansion are calculated, then the ratio of $m$th component to the $0$th component determines the strength of the $m$th mode. Another way to find the Fourier components, is to consider the position of 
the particles in the  disc  and expand the instantaneous distribution of disc particles as
\begin{equation}
A_m(t)=\sum_j \mu_j e^{im\phi_j}
\label{ex}
\end{equation}
in this case the amplitude $a_m$ is written as \citep[for example see][]{se16}
\begin{equation}
 a_m(t)=|\sum_j \mu_j e^{im\phi_j}|.
\end{equation}
where $\mu_j$ is the mass and $\phi_j$ is the cylindrical polar angle of each particle. We use this last method in this paper. These parameters are calculated at 
each time step to determine the amplitude of each mode as a function of time. So the bar strength, for which $m=2$, would be obtained by the ratio $\frac{a_2}{a_0}$.

The $m=2$ term of the expansion (\ref{ex}) gives the phase and amplitude of the bar as follows
\begin{equation}
A_2(t)=a_2 e^{2 i \phi(t)}
\label{pat}
\end{equation}
Therefore we can simply calculate the angular displacement of the bar, $\phi(t)$ in each time step. Furthermore, we calculate the time derivative, $\dot{\phi}(t)=\Delta\phi/ \Delta t$
in each time step. Finally we fit a tenth degree polynomial to the consecutive data points. We mention that $\dot{\phi}(t)$ is the pattern speed of the bar, i.e. 
$\Omega_p(t)=\dot{\phi}(t)$.

\subsection{Initial conditions}
\label{sui}
\begin{figure} 
 \centerline{\includegraphics[width=7cm]{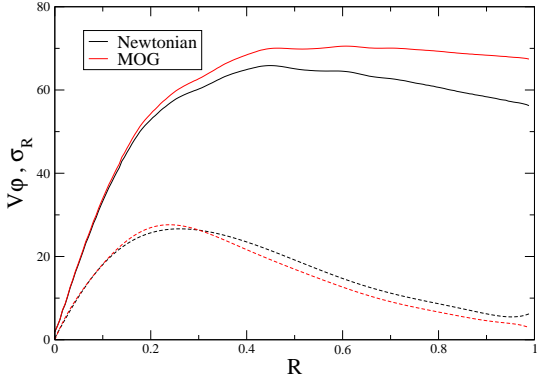}}
\caption[]{Initial rotational velocities for Newtonian and MOG exponential models. After about $R=0.1$, velocities in MOG gets higher than the Newtonian case. Dashed curves show the corresponding radial dispersion velocities. }
\label{vt0}
\end{figure}
We have modified the standard N-body code presented in \citet{aa1985,aa1994,aa2003} in order to study MOG effects on the disc dynamics. The original code assumes a 
Newtonian gravitational force between particles and uses a predictor-corrector method in which every particle is advanced in time with its own time step 
$\Delta t_{i}=(\eta f/\ddot{f})^{1/2}$, where $\eta$ is an accuracy parameter which affects the value of calculated individual time step of each particle,
considered to be $\eta=0.02$ in our calculations, and $f$, $\ddot{f}$ 
are the total force acting on the $i$th particle and its second derivative respectively. For more details we refer the reader to \citet{aa2003}.

Working in the cylindrical coordinate system $(r,\phi,z)$, we choose the initial surface density of particles to vary as $\Sigma\propto exp(-R/h)$ in a disc with initial radius 
$r_0$. Also we choose the units such that $G=1$, for simplicity we assume an equal mass for all particles and scale it to unity, i.e. $m=1$.  Also we adopt 
$r_0$ as our length unit. Therefore we use $R=r/r_0$ as the dimensionless length. In this case the initial radius of the disc is $R_0=1$. Albeit for ignoring the 
singularities that would show up from the close 
encounters of the particles during 
the numerical computations, the smoothing parameter $\epsilon$ should be included in the definition of the dimensionless length. Therefore in this paper by $R$ we 
mean
\begin{equation}
 R=\sqrt{\left(\frac{r}{r_{0}}\right)^{2}+\epsilon^{2} }
\end{equation}
In our main simulations, the value of $\epsilon$ is considered to be $\epsilon=0.05$.
Furthermore, we use a dimensionless time $\tau=t/t_o$ in which $t_0$ is a characteristic time defined in the item 4). Thus, using these units and 
equation (\ref{accmog}), the magnitude of the dimensionless force between two particles takes the form
\begin{equation}\label{f}
f(R)=\frac{1}{R^2}[1+\alpha-\alpha(1+\mu R)e^{-\mu R}]
\end{equation}
 and $\mu$ is a dimensionless parameter defined as $\mu=\mu_0 r_0$. We set the initial conditions using a similar procedure presented in 
\citet{op}. Let us briefly describe this procedure: 

1) In the case of Mestel-like disc, for initial positions we distribute $N$ particles on a disc of radius $R_0=1$. In order to set a Mestel-like surface density, we divide the disc into $N/10$ ring and 
into ten equally spaced slices ($\Delta \phi=36^{\circ}$). Then we place one particle in a random position within each cell. The resulting surface density is 
$\Sigma=N/2\pi R$.

On the other hand, in the case of the exponential disc $\Sigma\propto e^{-R/h}$ with the dimensionless scale length $h$, we have distributed $N$ particles on a disc
with initial radius $R_0=1$ using a different procedure presented in \citet{aa2003} (see chapter 8). We set $h$ to $0.2$ in our simulations. Changing this 
parameter would not change the main results. On the other hand, it is important mentioning that the typical scale length for spiral galaxies is $\sim 2 \text{kpc}$,
see \citet{bt}, and thus our choice is reasonable.

2) For adjusting the initial velocities, every particle is given an initial azimuthal velocity needed to balance with the centrifugal force and hold the particles in the 
circular orbits. For doing this, each particle is virtually rotated on a circle with radius equal to particle's distance from the centre of the disc and the quantity
$\sqrt{-R_{i}.~f_{i}}$ is computed every $1^{\circ}$. $f_{i}$ is the force acting on the given particle when it is settled at the position 
$r_{i}$. The average value of these 360 numbers is taken as the initial azimuthal velocity of each particle. The average value of this velocity for each ring is plotted 
in Fig. \ref{vt0} for both Newtonian and the MOGian disc.  As expected, the rotation curve in MOG is higher than the Newtonian disc.
\begin{figure*}
 \center
\centerline{\includegraphics[width=14.5cm]{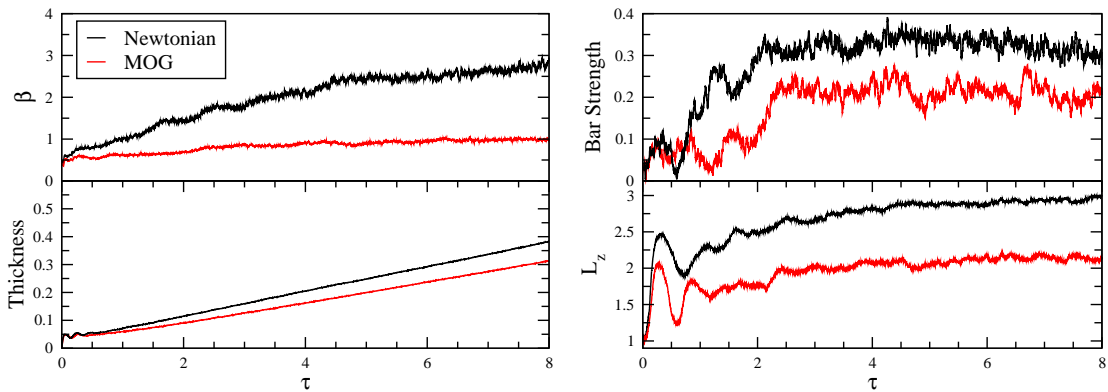}}
\caption[]{The top left panel shows the time evolution of the $\beta$ parameter in both models. The bottom left panel demonstrates the thickness of the discs with 
respect to time. 
The top right panel shows the bar amplitude in both models. The bottom right panel shows the vertical component of the total angular momentum in the outer part of 
the disc with inner radius $R=0.7$. }
\label{beta1}
\end{figure*} 

3) For making the disc stable against local perturbations at $\tau=0$, the velocity dispersions are added in three directions to each particle using the Toomre criterion
and the epicycle approximation. We recall that Toomre criterion is, \citep{toom64}
\begin{equation}
Q=\frac{\kappa\sigma_R}{3.36 G \Sigma}>1
\label{toomre}
\end{equation}
where $\kappa$ is the epicyclic frequency and $\sigma_R$ is the dispersion velocity in the radial direction. We can rewrite this criterion as $\sigma_R>\sigma_{\text{min}}$ where 
$\sigma_{\text{min}}$ is the minimum value of the radial dispersion velocity required for the local stability and is given by $\sigma_{\text{min}}=3.36 G \Sigma/\kappa$. On the 
other hand, from epicycle approximation we have a relation between the dispersion velocities as, see \citet{op} for more details
\begin{equation}
\frac{\sigma_{\phi}^{2}}{\sigma_{R}^{2}}\simeq \frac{1}{2}\left(1+\frac{d \ln~ v(R)}{d\ln R}\right)
\label{toom3}
\end{equation}
where $\sigma_{\phi}$ is the tangential velocity dispersion and $v(R)$ is the circular velocity. Therefore we may write
\begin{equation}\label{toom32}
\sigma_{\text{min}}\simeq  \frac{\sigma N\Delta}{v(R) \sqrt{1+\frac{d\ln v(R)}{d\ln R}}} 
\end{equation}
\begin{equation}\label{toom33}
\sigma_{\phi}\simeq \frac{\sigma}{\sqrt{2}} \frac{N\Delta}{v(R)} 
\end{equation}
where $\Delta=1$ for the Mestel disc and for the exponential disc we have
\begin{equation}
\Delta=\frac{R\,e^{-R/h}}{h(h-(1+h)e^{-1/h})}
\end{equation}
also $\sigma=0.378$. Setting the initial $\sigma_{R}$ to be 20\% larger than the minimum value, we have $\sigma=0.454$. 
We use this parameter, i.e. $\sigma$, in 
order to control the response of the system to local perturbations.

One should note that the Toomre criterion in MOG is different from (\ref{toomre}). In fact generalised version of this criterion in MOG can be written as 
$\sigma_R>\sigma_{\text{MOG}}$ where $\sigma_{\text{MOG}}>\sigma_{\text{min}}$, for more details see \citet{ro2015}. However because of the above mentioned 20\% 
enhancement, the difference between the local stability criteria in MOG and Newtonian gravity does not matter, and consequently both models are locally stable. Also 
the thickness of the disc in the $z$ direction can be taken into account by adding a velocity dispersion in the vertical direction as $\sigma_{z}=\sigma_{\phi}$.

4) Finally, after adding the velocity dispersions, we should re-establish the equilibrium between the gravitational and the centrifugal forces. To do so, the average
velocity $v$ of each ring is multiplied by a factor to make it equal with the average velocity of ring before adding the velocity dispersions.

Indeed, the time evolution of the disc is described in terms of the initial average rotation period $t_0$ of the outer ring and the disc is initially rotating in the counterclockwise direction. 
The Number of particles is set to $N=2500$ for the main simulation and has been increased to $10000$ in order to check the effects of number of particles, $N$, on the results. In fact, although our results are also unaltered for particle numbers over $10000$, we have not included them, since the error in relative energy conservation gets higher than $1$\%

It is necessary to mention that the code computes direct summation of the forces over all particle pairs. We have used this method since modifying the gravitational force in the code is straightforward in this case. But the computation time scales as $N^2$ and makes the code inefficient for large $N$. Therefore we restrict ourselves to $N \leq 10000$. However we believe that although $N$ is small, it is enough to fulfil our simple purpose in this paper. In other words, we are interested on the overall effects of MOG on the global stability of self-gravitating discs and not on the details of its time evolution. Consequently it does not seem necessary to use large $N$ in the simulation. For a paper with a similar aim in which a similar code is used we refer to \citet{chris} where $N\leq 1200$. Of course to have a more precise view on the evolution of a galactic disc in MOG, it is necessary to employ higher number of particles in the codes which use different methods for computing the gravitational force \citet{gh}.

It is helpful here to reiterate that \citet{op} used simulations with only $N=150$ to $500$ and for the first time showed that rigid halos have stabilizing effects on the self-gravitating discs. Their result has been not ruled out in simulations containing rigid halos and large number of particles $N$. From this perspective, and considering the limitations of the code and keeping in mind the main aim of the paper,  we have restricted ourselves to $N \leq 10000$. However, in section \ref{T} we have  applied three tests to the code in order to check the reliability of the results. More specifically, in the case of Mestel-like disc we present an analytic description which confirms the results obtained from our simulations.

The initial rotational velocity of both Newtonian and MOG models is plotted in Fig. \ref{vt0}. As it is expected, in large distances from the centre the rotation 
curve in MOGian disc is almost flat and tangibly higher than the Newtonian disc. On the other hand, until about $R=0.1$, the velocities in both models are 
equal. In fact in real spiral galaxies MOG modifications to the gravitational law give larger accelerations to the stars and so increases their rotational velocity. 
Furthermore, $\sigma_R$ for MOG model is larger than the Newtonian case 
for $0.1\lesssim R\lesssim 0.31$ and after $R\sim 0.31$ gets smaller. In fact a same value for the Toomre parameter has been used 
in both models therefore we can write $(\sigma_R \kappa)_{\text{N}}=(\sigma_R \kappa)_{\text{MOG}}$. On the other hand, one may show that epicycle frequency in MOG is
smaller than the Newtonian model in the interval $0.1\lesssim R\lesssim 0.31$ and then gets larger. Consequently one may simply explain the behaviour of the radial velocity dispersion in both 
theories. Albeit one may expect that the epicycle frequency in MOG to be larger than the Newtonian case in any radius since MOG leads to stronger gravitational 
attraction and consequently increases the epicycle frequency, i.e. $\kappa\propto \sqrt{F(R)}$, where $F(R)$ is the magnitude of the gravitational force. However, 
one should note that the first derivative of the force with respect to radius is also important for magnitude of $\kappa$ as 
\begin{equation}
\kappa(R)=\sqrt{\frac{dF}{dR}+\frac{3}{R}F(R)}
\end{equation}
Therefore increasing the strength of the force does not necessarily mean that $\kappa$ has been increased. In the case of an infinite Mestel disc we have found the exact form of the rotation curve, 
see equation \eqref{sk1} in the appendix. One may easily use equation \eqref{sk1} and compare $\kappa$ in both theories.

It is necessary to mention that in order to determine the 
dynamical properties of our models, we need a well-defined
centre among the particles about which to perform the simulations. The centre of our N-body system wanders
from the centre of our coordinate system. In fact the centre of mass of the system in principle has a non-zero initial velocity. Consequently the centre will move
during the 
simulation and may cause non-real effects and numeric artefacts. We use a simple procedure to find the centroid of the system. For each time step, we calculate the
Fourier 
amplitudes with respect to an origin $(x_0,y_0)$ initially chosen near the main origin of the coordinate system, and change this point smoothly until find a point for 
which the 
amplitude of the first Fourier mode $m=1$ is minimum. The final centroid $(x_0,y_0)$ has been shown by a red plus in Fig. \ref{fig:evolution} for different $\tau$.

As a final remark, the change in total energy can be monitored at regular time intervals in order to measure the global error. In other words, we measure the 
relative energy 
conservation $|\Delta E|/E_0$ in each time step. Where $E$ is the total energy of the particles and $E_0$ is its initial value at $\tau=0$. In our main simulations
this quantity is always lower than 1\%.

\section{\small{Results}}
\label{res}
\subsection{Stability parameter $\beta$}

As mentioned above, one useful criterion to check the stability of the discs is the time evolution of $\beta$ parameter. This parameter measures the ratio of the 
random motions to 
rotational motions. \citet{op} showed that for a bare disc, this parameter rapidly grows and the disc evolves to a stable pressure dominated configuration, i.e. the 
stellar bar. 
On the other hand, $\beta$ does not grow rapidly in the presence of a rigid halo. As we mentioned before, this fact has been approved in several numerical studies. 
Regarding our
purpose in this paper, it is important to see how fast the disc is evolved to the hot stable state. The time evolution of this parameter in both MOG and Newtonian 
models has been 
shown in top left panel of Fig. \ref{beta1}. The initial value of $\beta$ for MOG disc is slightly smaller than the Newtonian case. This fact is expected since MOG 
increases the rotational velocity of the particles. 
\begin{figure} 
\centerline{\includegraphics[width=6.75cm]{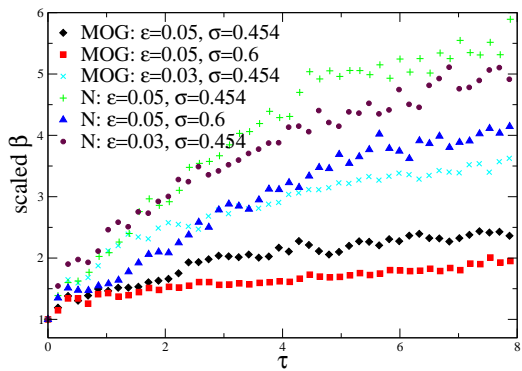}}
\caption[]{Stability parameter $\beta$ for different values of $\epsilon$ and $\sigma$. $\beta$ is scaled to its initial value at $\tau=0$. Therefore all curves 
start from $1$.}
\label{beta2}
\end{figure}

The $\beta$ parameter in Newtonian disc, as expected, grows rapidly and reaches $\beta\sim 1$ at $\tau\sim 1$, i.e. at the first rotation period. This means that random motions 
rapidly dominate the disc against the rotational motions. Rapid growth continues until $\tau \sim 4.5$ and after that $\beta$ grows with
a smaller and almost constant slope.

For the MOG disc, it is clear that $\beta$ parameter raises much slower. It grows to about $0.75$ at $\tau=2.5$, then it grows with an almost constant slope and 
reaches to about $1.$ at the end of the simulation.
This means that unlike the Newtonian case, the disc remains approximately rotationally dominated and the bar formation happens with a very small rate compared to the Newtonian disc. 

Another quantity for measuring the departures from the initial disc shape, is the thickness of the disc. One may use the root-mean-square value of $z$ of the particles as the 
thickness of the disc. The time evolution of the thickness has been shown in the bottom right panel of Fig. \ref{beta1}. In Newtonian disc the thickness increases 
rapidly with time. As mentioned in \citet{op}, this fact is in agreement with the rapid bar formation. Thickness in the MOG model increases  slower.

In order to show that our main results are not sensitive to the magnitude of the softening parameter $\epsilon$, we have performed simulations for different values of 
$\epsilon$. To see
the difference, we have plotted the scaled $\beta$ parameter, i.e. $\beta/\beta_0$ where $\beta_0$ is the initial value of $\beta$, in Fig. \ref{beta2}. In the case of 
Newtonian disc, 
by comparing the graph of circles ($\epsilon=0.03$) and pluses ($\epsilon=0.05$) it is clear that the evolution of the $\beta$ parameter does not change significantly by 
changing $\epsilon$. In MOG disc, by comparing the diamonds ($\epsilon=0.05$) and crosses ($\epsilon=0.03$), 
it seems that $\beta$ parameter in MOG is more sensitive to changes in $\epsilon$ than in Newtonian gravity.

\begin{figure*}
 \center
  \includegraphics[width=\textwidth]{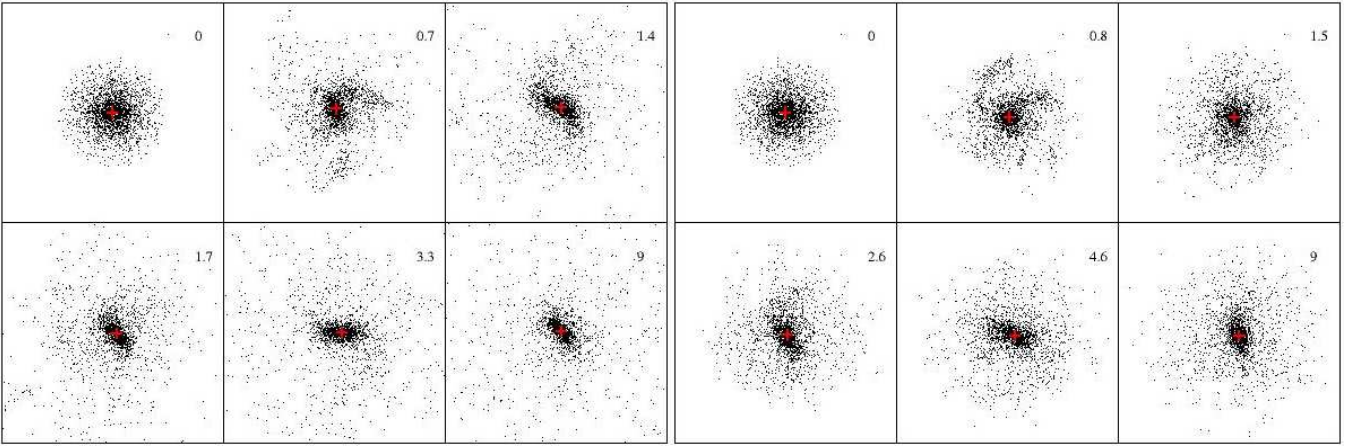}
  \caption{The evolution of the disc with respect to time projected on the $x$-$y$ plane. The left and right panels belong to the  Newtoninan and MOG model respectively. Both models are exponential, i.e. $\Sigma\propto e^{-R/h}$.} 
  \label{fig:evolution}
\end{figure*}
\subsection{Bar growth}
In this subsection we discuss the bar growth in both models. Let us start with the Newtonian disc. The left panel in Fig. \ref{fig:evolution} shows the time evolution
of the Newtonian disc, projected in the equatorial plane. The top left block shows the initial disc. A weak bar starts to form rapidly. The $m=3$ mode is apparent at $\tau\simeq 0.7$ and after 
 $\tau \simeq 1$, the bar grows stronger and the spiral arms appear at $\tau\simeq 1.4$. They fade and reappear until $\tau\simeq 3.3$ but the bar is a permanent 
 feature and stays almost unchanged until the end of the simulation.

In the top right panel in Fig. \ref{beta1}, we have plotted the bar amplitude $a_2$. The black curve corresponds to the Newtonian disc. 
In this case, after experiencing a temporary minimum, the bar amplitude rapidly increases and reaches to $0.27$ at $1.2$. Then after a temporary decrease, raises again
about $0.35$ at $\tau=2$. Then the disc settles in an almost equilibrium state, i.e. the bar. However it is clear that the bar strength decreases with a small 
slope after $\tau\simeq 5$. On the other hand, as we mentioned, the stability
parameter $\beta$ increases smoothly. This means that although the bar is weakened, the random motions are increasing. Therefore one may conclude that the bar decays
slowly to a spherical pressure dominated configuration, say a stellar bulge. We checked the existence of this spherical configuration by projecting the positions of 
the particles in the $z-y$ and $z-x$ planes. Furthermore, we will see in subsection \ref{othermodes} that the presence of other modes is tangible compared to 
the $m=2$ mode in the first rotation period, specially for $m=3$. The bar starts to be the dominant mode 
after $\tau\simeq 1$ and reaches its maximum strength at about $\tau\simeq 2$.

The red curve in the top right panel in Fig. \ref{beta1} is the bar amplitude in MOG model. In this case the growth rate is much less than the Newtonian case. More 
specifically, until $\tau=2.5$ the $m=2$ mode forms and decays frequently and the bar strength is weak. In other words, there is an explicit oscillation in the bar 
amplitude during the first $2.5$ rotation periods.
It is shown in section \ref{mesteldisk} that these oscillations exist also in the Mestel-like disc. We emphasise that the
existence of transient spiral patterns is not the case in the Newtonian disc (without halo). However it seems that they are a key feature in the MOG models.
These oscillations in the strength might be an indication of the existence of several modes which are beating.

After $\tau=2.5$, the 
disc settles into an equilibrium configuration,
with bar amplitude relatively constant thereafter, $a_2/a_0\sim 0.2$. This fact is also different from the Newtonian disc. In fact, constancy of the bar strength 
shows that, unlike the Newtonian case the bar mode is a stable configuration and does not decay to a spherically symmetric system. In other words, the final state of
the MOG disc, is a shortened bar rotating in the plane of the disc. 

The transient nature of the spiral patterns in the MOG model signals that the propagation of the stellar density waves and their interaction with the relevant 
resonances in the context of MOG could be different from the Newtonian gravity. This issue needs more careful studies in order to determine the origin of this 
interesting feature. However, it is interesting to mention that there are observations which verify that spiral patterns should be short-lived rather than long-lived
structures, see \citet{se11}.

\begin{figure} 
\centerline{\includegraphics[width=8.5 cm]{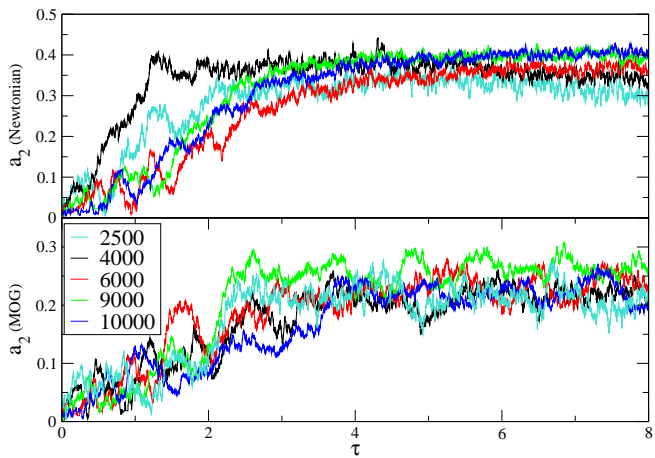}}
\caption[]{Time evolution of the bar strength for different number of particles $N$ in the disc. For both Newtonian and MOG discs, the final values of the bar strength is somehow independent of $N$.}
\label{fig:bardiffN}
\end{figure}

In order to check the dependency of the results to $N$, we have shown the bar strength for some different values of $N$ in Fig. \ref{fig:bardiffN}. It illustrates that the bar amplitude evolves in a similar trend for all $N$. In fact in both Newtonian and MOG discs, the final value of the bar amplitude is somehow independent of $N$. More specifically, the time averaged value of the bar magnitude in both theories and for different particle numbers is presented in Table \ref{tab1}. The fractional difference $\delta$ is defined as
\begin{equation}
\delta=\frac{\left< a_2\right>_{\text{N}} - \left< a_2 \right>_{\text{M}}}{\left< a_2\right>_{N}}
\end{equation}
where subscript "N" and "M" stand for Newtonian and MOG models respectively, and $N$ should not be confused with particle numbers. This parameter shows that the averaged value of the bar strength in Newtonian cases is higher than the MOG models. Although there seems to be some random change in $\Delta$ according to different particle numbers, the main feature that the bar strength in MOG is smaller than the Newtonian case is unchanged. The mean value for the averaged bar 
magnitude in Newtonian is 0.29 while this parameter is 0.17 for MOG, which results in a fractional difference $\bar{\Delta} \simeq 41.38\%$. Related discussions of the Mestel-like disc will be presented in section \ref{mesteldisk}.

\begin{table}
\caption{Time averaged value of the bar strength over $\tau=0$ to $\tau=8$ for MOG and Newtonian models, in both exponential and Mestel-like discs with different number of particles. \label{tab1}}
\begin{center}
\begin{tabular}{ccccc}
\hline
Model & $N$ & $\left<a_2\right>_{\text{M}}$ & $\left<a_2\right>_{\text{N}}$ & $\delta\times100$ \\ 
\hline
\hline
	& 1600	& 0.16	&0.25	& 36.0	\\ 
 	& 2500 	&0.17	&0.28	& 39.28 \\
 	& 4000	&0.17	&0.33	& 48.48	\\
 $\Sigma\propto \exp (-\frac{R}{h})$	& 6000	&0.18	&0.26	& 30.77 \\
  	& 7000	&0.18	&0.29	& 37.93	 \\
 	& 8000	&0.10	&0.28	& 64.29 \\
 	& 9000	&0.20	&0.30	& 33.33 \\
 	& 10000	&0.17	&0.30	& 43.33 \\
 	 \hline
 Mean values	& -	&0.17	&0.29	& 41.38	\\
 \hline
	& 1600	&0.17	&0.25	& 32.0	 \\ 
     & 2500	&0.17	&0.29   & 41.38 \\
 	& 4000	&0.24	&0.28	& 14.28 \\
   $\Sigma\propto \frac{1}{R}$	& 6000	&0.24	&0.30	& 20.0\\
	& 7000	&0.18	&0.29	& 37.93 \\
 	& 8000	&0.23	&0.30	& 23.33 \\
 	& 9000	&0.24	&0.31	& 22.59 \\
 	& 10000	&0.19	&0.29	& 34.49\\ 
  \hline	
  Mean values	& -	&0.21	&0.29	& 27.59	\\
 	 \hline
\end{tabular}
 \end{center}
 \end{table}

\begin{figure}
\includegraphics[scale=0.83]{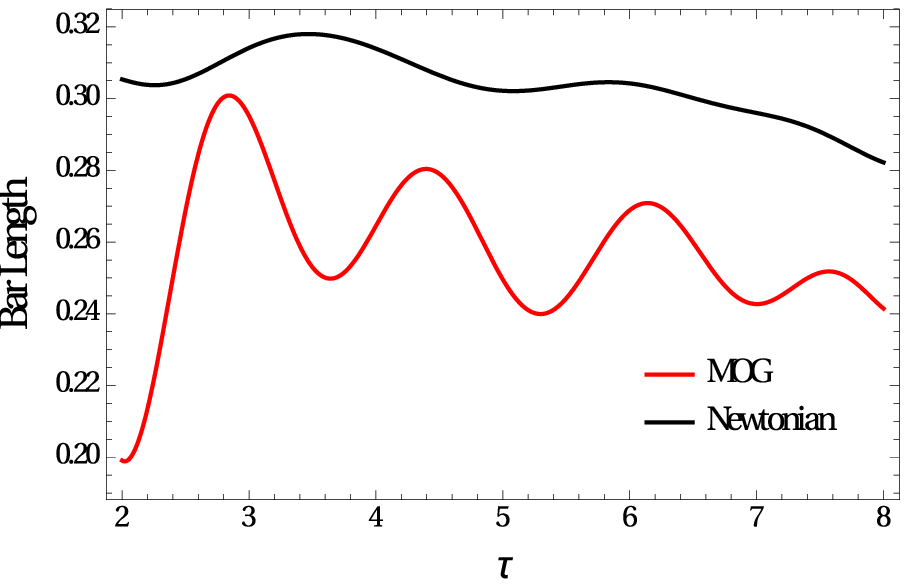}
\hspace*{.1cm}
\includegraphics[scale=0.81]{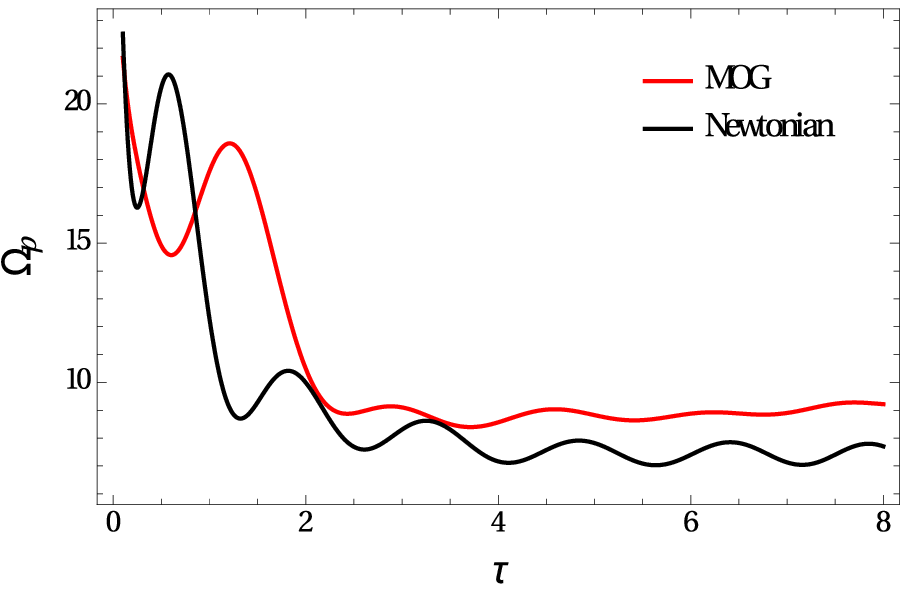}\caption{Top panel: the time evolution of the bar length in both exponential models. Since in both models, the bar is fully formed after $\tau\sim2$, we have plotted its length for $\tau>2$. Bottom Panel: evolution of the pattern speed with time. The black curve belongs to the Newtonian disc and the red curve is for the MOG model. For very small times, these curves are not reliable since the pattern has been not fully formed at these times.}\label{newfig}
\label{fig:barpat}
\end{figure}

It is also interesting to compare the length of the stellar bar in both theories. We have plotted the length of the bar, after fitting a nineteen degree polynomial, in Fig. (\ref{fig:barpat}, top panel) for both 
models. To do so, at each time $\tau$ we assume a rectangle with width $\sim 0.2$ and length $\sim 1$ around the line $y=\tan \phi(\tau)x$, where $\phi(\tau)$ is the 
angular displacement of the bar. Then after dividing this rectangle to small elements, the bar length is chosen to be the length at which the number density of the 
particles is less than $30$\% of the centre number density. From Fig.(\ref{fig:barpat}, top panel), it is apparent that the bar length is longer in the Newtonian model. It grows 
to about $0.32$ at $\tau=3.4$ and then declines slowly. This behaviour is consistent with the bar strength. On the other hand, as expected, the bar length in MOG is 
shorter. Furthermore, there are obvious oscillations in the bar length in MOG. These oscillations are reminiscent of a similar behaviour in the N-body simulations of 
the spiral discs in the context of MOND, see Fig. 8 in \citet{ti}.

 The main result of this section is that the final bar strength in MOG is weaker and the growth rate is smaller than the Newtonian case. This result is satisfactory and is what one may expect from a modified theory of gravity. In other words, as we already mentioned, modified gravity 
has to play a same role as the dark matter halo for suppressing the bar growth rate. However this result seems somehow puzzling. In fact there is no difference between
MOG and Newtonian model in the inner disc, and the differences appear at large distances from the centre. On the other hand, the bar is formed in the inner disc. 
Therefore one may expect that its dynamics would be similar to the Newtonian case. However, as our simulations show, the dynamics of the bar is significantly
different in these theories.

\subsection{Pattern speed $\Omega_p$}

We also calculate the pattern speed $\Omega_p(t)$ using equation (\ref{pat}). The result has been illustrated in Fig.(\ref{fig:barpat}, bottom panel) after fitting a
nineteenth degree polynomial. Since the bar rotates smoothly, $\Omega_p(t)$ 
is not as noisy as other functions that we have already plotted. We assume that the pattern exists even in small times. The black curve corresponds to the Newtonian model. In this
case the pattern speed rapidly decreases until $\tau\sim 0.25$ and then grows and experience a maximum at $\tau\sim 0.5$. There is another maximum 
at $\tau\sim 1.8$. After this time, $\Omega_p(t)$ decreases until $\tau=2.2$ and gets almost constant, i.e. $\Omega_p\sim 8$,  until the end of simulations. Albeit 
there are some small amplitude oscillations.

It is evident that the pattern speed in the MOG model is slightly larger than the Newtonian disc. One may expect this fact since the initial circular velocity of the particles are 
larger in the MOG model. In this case as in Newtonian model, $\Omega_p$ first starts with a decreasing phase. However at around $\tau\sim 0.5$ starts to raise and after 
experiencing a maximum at $\tau\sim 1.3$ starts to decrease and finally reaches the almost constant value $\Omega_p\sim 9$. Existence of maximums in the evolution of the 
pattern speeds of both models can be simply related to the evolution of the stellar bar. It is clear form Fig. \ref{beta1} that in the same period of time that the pattern speeds 
experience a maximum, the bar magnitudes experience a minimum. This means that the stellar bar is temporarily converts its shape to an almost spherical system. Therefore the 
characteristic size of the pattern gets smaller. Naturally one may expect an increase in the angular momentum of the pattern in this situation. Therefore it is expected that the 
pattern speed grows in this period of time.

\begin{figure}
\centerline{\includegraphics[width=8.7cm]{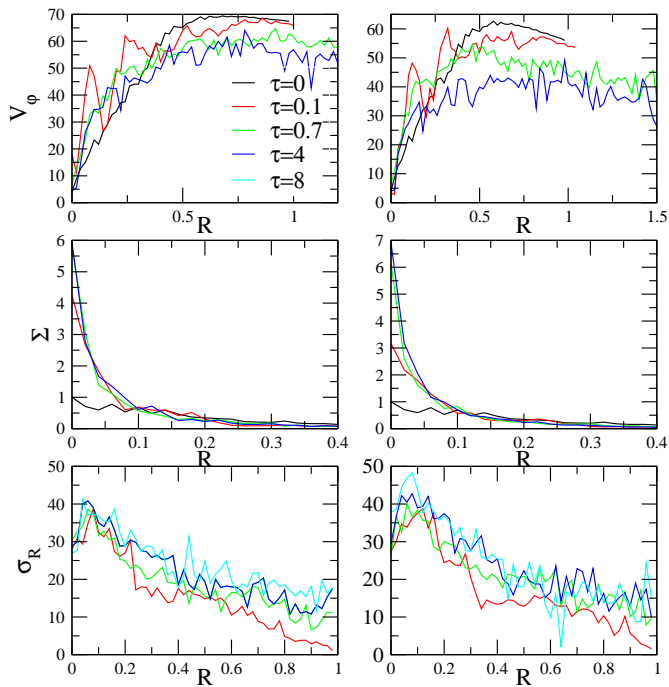}}
\caption[]{Radial dependency of azimuthal velocity, scaled surface density, and radial velocity dispersion for MOG (left panels) and Newtonian (right panels) at four 
different times.}
\label{radial}
\end{figure}

\subsection{radial profiles of $\Sigma$, $v_{\phi}$ and $\sigma_r$}
It is also instructive to study the radial behaviour of the main quantities like surface density $\Sigma$, radial velocity dispersion $\sigma_r$ and the circular 
velocity $v_{\phi}$. We have plotted the radial profile of these quantities at different times in Fig. \ref{radial}. The left and right panels belong to MOG and 
Newtonian exponential discs respectively. It is clear from top panels that although the rotation curve in both models change with time, the changes are smaller in 
the MOG model. Moreover, we found that $v_{\phi}$ radial profile remains almost unchanged after $\tau \sim 0.4$ in Newtonian model and $\tau\sim 0.7$ in the MOGian 
disc.

The projected surface density profiles have been plotted in the middle panels. We have scaled $\Sigma$ to its initial value at $R=0$. As expected, surface 
density grows at $R\sim 0$ in both models, and this growth is higher in the Newtonian disc.

In the case of radial dispersion velocities, in the bottom panels, it is clear that in small radii $\sigma_r$ grows faster in the Newtonian case. More specifically,
at $\tau=8$ the maximum value of $\sigma_r$ reaches to $48$ in the Newtonian disc while it is 15\% smaller in the MOGian case.

\subsection{Angular momentum exchange}
Angular momentum exchange between galactic components is a key factor in the bar formation. Angular momentum exchange plays also a central role for describing many accretion 
phenomena in astrophysical systems, for example see \citet{ss} for viscosity driven angular momentum exchange in accretion discs. In our models, in order to buckle in
the centre as a bar, the disc needs to loose angular momentum. For example \citet{ti} and \citet{se16} has studied the angular momentum transfer
from the disc to the surrounding halo during the stellar bar formation. \citet{se16} showed that angular momentum transfer from the galactic disc to the live halo, may behave 
like a source for the bar growth. 

Although in our models there is no halo, the angular momentum can be transported throughout the disc by the torques that the stellar bar exerts on the outer parts of the disc. 
Therefore we checked the angular momentum transfer from inner parts of the disc, to the outer parts. To do so, we have calculated the vertical component of the angular momentum, 
i.e. $L_z$, with time for a part of the disc outside the radius $R=0.7$. The result has been shown in the bottom right panel in Fig. \ref{beta1}. The black and red curves 
correspond to the Newtonian and MOG models respectively. The angular momentum transport in the Newtonian disc happens with a larger rate. This is expected since the origin for 
this exchange is the stellar bar, and the bar is stronger in the Newtonian disc and also this bar grows with a faster rate. 
This difference between the two models can also be obtained using analytic descriptions in the case of Mestel-like disc. The 
related discussions are included in section \ref{mesteldisk}, and calculations are presented in the appendix. 

It is interesting to mention that because of the lower azimuthal velocity at larger radii in the Newtonian model, one may expect an opposite behaviour. However, we plotted the surface density of the discs at larger radii in both models, and realized that the surface density of Newtonian Model, is larger than the MOG model at $r\gtrsim 1.1$. This result holds for different times in the evolution of the disc, and means that the spread of the particles to larger distances is higher in the Newtonian model.

\subsection{$m\neq 2$ modes}
\label{othermodes}
\begin{figure}
\centerline{\includegraphics[width=8cm]{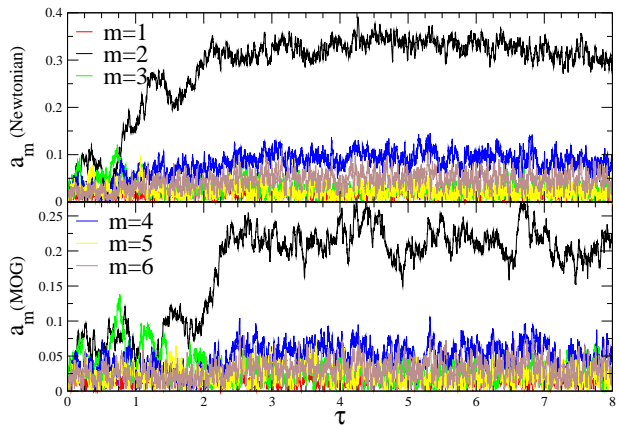}}
\caption[]{Time evolution of different Fourier modes in a exponential disc for Newtonian and MOG models. The dominant mode for both models is $m=2$. It is clear that even modes are stronger than the odd modes. The $m=3$ mode in MOG models is stronger than the Newtonian case.}
\label{fig:2in1}
\end{figure}

Although from Fig. \ref{fig:evolution} it is somehow evident that other modes ($m\neq 2$) are not strong enough to dominate the dynamics, we have shown their amplitude, i.e. 
$a_m/a_0$, in Fig. \ref{fig:2in1}. The top panel corresponds to the Newtonian disc and the bottom panel is for the MOG model. In general even modes are stronger than 
the odd modes. However the dominant mode in both models is the bar mode. In both models, around $\tau\sim 0.7$ there is an explicit enhancement in the amplitude of 
$m=3$ mode. In the MOG model this mode even dominates the bar mode for a short period of time. Furthermore, at $\tau\sim 1$ the amplitude of $m=4$ mode reaches its 
maximum. 
It is important to mention that the oscillatory nature of the modes in the MOG is not restricted only to the bar mode and appears for other modes.

\begin{figure}
 \centerline{\includegraphics[width=8cm]{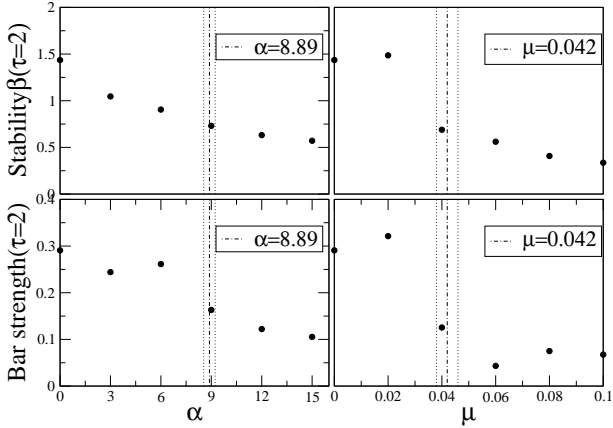}}
\caption[]{Top left panel shows the $\beta$ parameter at $\tau=2$ for a constant $\mu$ and different values of $\alpha$. In the top right panel $\alpha=8.89$ and $\beta$ has been computed for different $\mu$. In the bottom panel, with a same procedure, the bar magnitude has been measured at $\tau=0.5$ for different values of $\alpha$ and $\beta$.}
\label{fig:diffAlphaMuBar}
\end{figure}

\begin{figure}
 \centerline{\includegraphics[width=7.8cm]{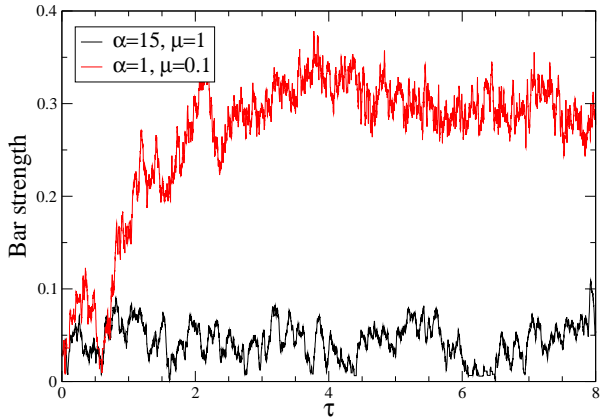}}
\caption[]{Effect of MOG free parameters on the bar evolution. $\mu$ is shown with its scaled value. It is clear that large parameters can totally prevent the bar formation.}
\label{fig:MOG}
\end{figure}
 \subsection{Effects of MOG's free parameters on the bar growth}

 \begin{figure*}
 \center
  \includegraphics[width=13cm]{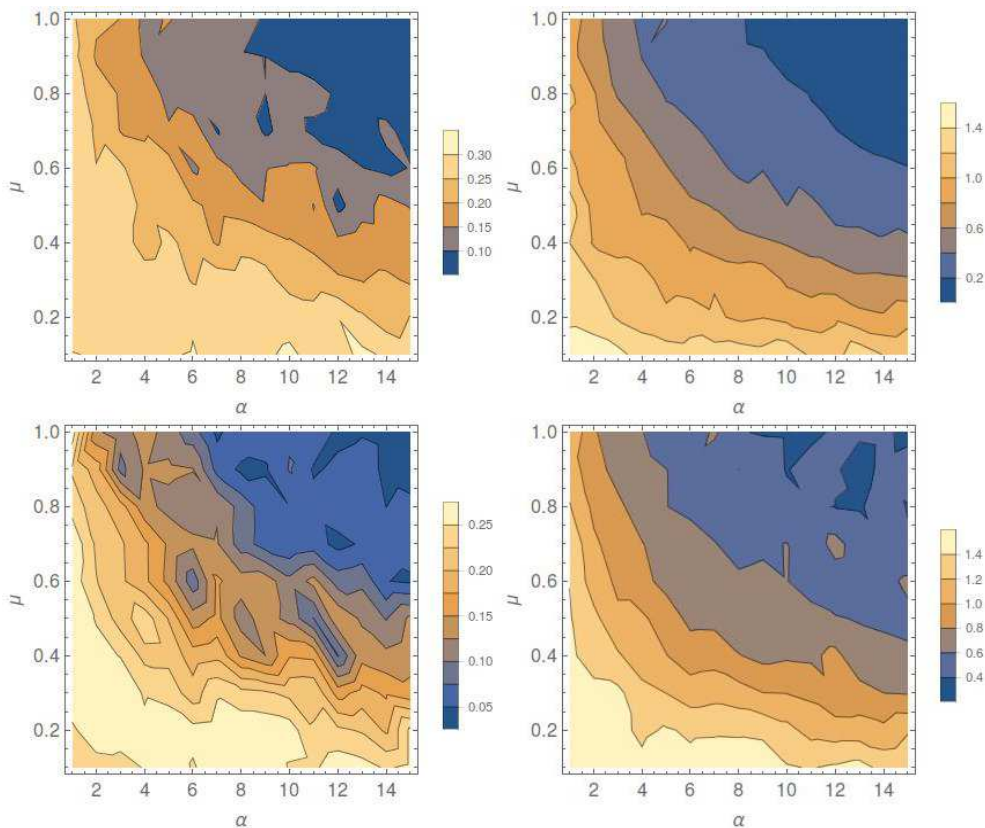}
  \caption{The time averaged vale of $\beta$ has been indicated in the left panel as a contour plot for different values of $\alpha$ and $\mu$. Each point on these panels means a separate simulation, each panel has 150 points. In the right panel the averaged value of the bar amplitude has been computed for different free parameters. Both panels approve that the bar instability is suppressed by increasing the free parameters. }
  \label{contur}
\end{figure*}

 We recall again that MOG possesses two free parameters, $\alpha$ and $\mu_0$, in the weak field limit. In principle, these parameters are not universal and they may 
differ from scale to scale \citep{hagh}. 
As we already mentioned, in the case of spiral galaxies, the current observational values for these parameters are $\mu_0=0.042\pm 0.004~\text{kpc}^{-1}$ and
$\alpha=8.89\pm 0.34$. In our simulations
we have set $\alpha=8.89$ and $\mu_0=0.042~\text{kpc}^{-1}$. In this section we study the response of the stellar bar to the changes in these free parameters. 

To do so, we check the stability parameter $\beta$ for different values of $\mu$ and $\alpha$. The result has been illustrated in the top panels of Fig. 
\ref{fig:diffAlphaMuBar}. 
In this figure the observational range of the parameters has been indicated by dot dashed
lines. In the top left panel, $\mu$ is constant and is set to $0.42$ (in its
dimensionless form), and the $\alpha$ parameter is varied. The stability parameter has been computed at $\tau=2$ for different values of $\alpha$. It is evident that
by increasing this 
parameter $\beta$ decreases. In other words, by increasing $\alpha$ one may stabilise the disc. At first sight, this result seems somehow bizarre since larger $\alpha$
leads to 
stronger gravitational force. Consequently one may expect that a stronger gravitational force would destabilise the disc. However we recall that existence of a halo, 
also strengthens the radial gravitational force while suppressing the global gravitational instability. 

In the top right panel in Fig. \ref{fig:diffAlphaMuBar}, we set the $\alpha$ parameter to $8.89$ and check the response of the disc to different $\mu$. As in the 
$\alpha$ parameter case, $\mu$ has also stabilising effects on the disc. 

Also we have done the same procedure for the bar amplitude and measured it at $\tau=2$
for different values of the free parameters. The result has been plotted in 
the bottom panels of Fig. \ref{fig:diffAlphaMuBar}. This result also approve the above mentioned point that both free parameters have stabilising effects on the disc More 
specifically, large values of free parameters can totally suppress the bar formation, for example see Fig. \ref{fig:MOG} in which we have compared the bar evolution
for small and large values of free parameters. This fact raises the 
question that if MOG can explain the existence of unbarred galaxies. It is worth mentioning that although the dark matter halo provides a reasonable and satisfactory 
picture for the bar
evolution, it leaves also some unanswered questions. In some cases a halo can even speed up the bar growth, for example see \citet{at2002} and \citet{saha}. Recently 
\citet{se16} using N-body simulations found out that angular momentum transfer between a live dark matter halo and the stellar disc can effectively trigger the bar instability.
This means that if disc galaxies are assumed to embed in live dark matter halos, then the observational evidence that more than 35\% of them lack a strong bar is still a serious challenge.

On the other hand, as we already mentioned, MOG free parameters are not universal constants and have different values in different 
environments. For example, in Table 1 of \citet{m2013}, these parameters are fitted using observational data, and it is shown that for some of the LSB galaxies in their sample, are large and comparable to the values in Fig. \ref{fig:MOG} for black curve. In other words, in principle, MOG may explain the unbarred galaxies provided that their characteristic values for $\alpha$ and $\mu$ are large enough. Therefore, it seems that the special case of unbarred galaxies should be 
studied more carefully. In other words, it is crucial to find the observational value of the free parameters in these systems and then study their global stability using N-body simulations. We leave this issue as a future study.

In order to find a more reliable method to investigate the effects of these parameters on the evolution of the stellar bar, we have performed several simulations and 
computed the 
time average value of $\beta$ and $a_2$ during the entire simulation (five rotation periods). The results of $\beta$ and $a_2$ have been illustrated in left 
and right panels of the contour plot Fig. \ref{contur} respectively. The upper panels belong to the exponential disc. In the lower panels we have shown the 
corresponding results for the Mestel-like disc. This figure provides a full
parameter space defined by $0\leq\alpha\leq 15$ and $0\leq\mu\leq 1$. In fact we have performed $150$ simulations , for each model, to cover the parameter space by a symmetric mesh. Albeit by 
increasing the mesh points and consequently the number of simulation, one may find a better resolution. Contour curves indicate points $(\alpha,\mu)$ for which the 
measured quantity, i.e. $\large<\beta\large>$ in the right panel and $\left<a_2\right>$ in the left panel, are the same. This figure completely approves
the results presented 
in Fig. \ref{fig:diffAlphaMuBar}. The left panels in Fig. \ref{contur} shows that by increasing the free parameters the bar instability is suppressed more. 

The right panels show the averaged bar amplitude. Although they are not as smooth as the left panels, specially in the exponential disc, they also show that 
increasing the free parameters will stabilise the disc. It is clear that the oscillatory nature of the bar amplitude in MOG also appears in this parameter space.

\section{Tests}
\label{T}
The limited number of particles in the code is a major concern which made us perform some tests on the reliability of the code. 
First we applied the code to a Mestel-like disc. Fortunately, in this case we can derive some of the results analytically. As we will show in the next subsection 
the analytical description confirms our numeric results. As another test, we re-derive the stability criterion introduced by \citet{ef} with a different numerical 
method. And to diminish the possible particle noise, we check our results using a quiet start \citep{se89}.

\begin{figure*}
 \center
  \includegraphics[width=\textwidth]{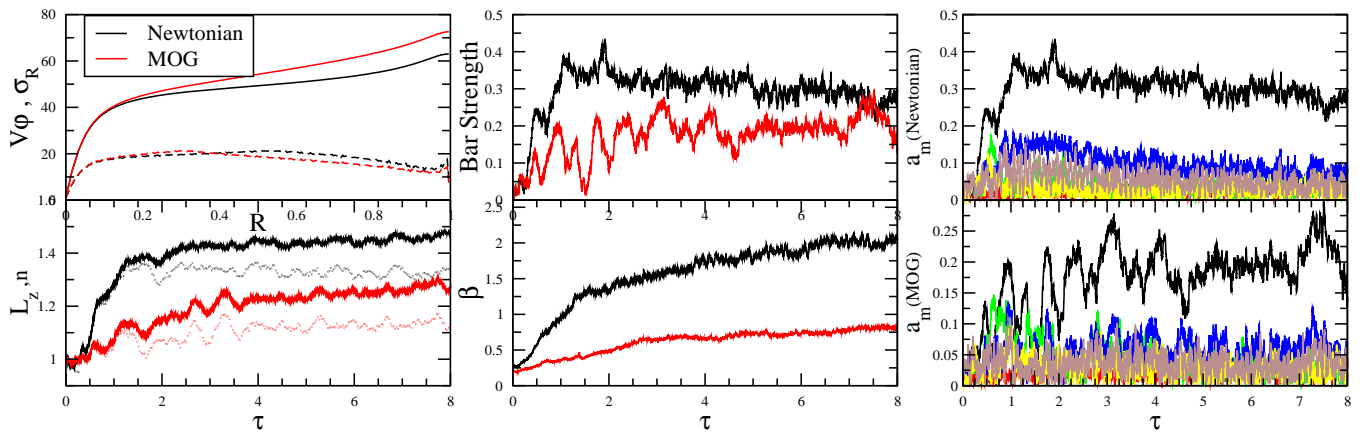}
  \caption{Results for the Mestel-like disc, i.e. $\Sigma\propto \frac{1}{R}$: The top left panel shows the circular velocity in Newtonian gravity and MOG. 
  Of course, this is a toy model and the rotation curve is not consistent with the observations. The dashed curves in this panel are the radial velocity dispersions 
  in both models. The bottom left panel is the vertical component of the total angular momentum of a ring with inner radius $R = 0.7$ in both models. The fractional 
  number of particles,i.e. n, in the outer part of this radius has been indicated by dotted black curve in the Newtonian case and by the dotted red curve for the MOG 
  model. The top middle panel is the bar strength in Newtonian gravity and MOG. The bottom middle panel is the $\beta$ instability parameter and the right panel is 
  the magnitude of the other Fourier modes, in the same order as in Fig.\ref{fig:2in1}.} 
  \label{mestel1}
\end{figure*}

\subsection{\small{Global stability of a Mestel-like disc in MOG }}
\label{mesteldisk}
The transfer of angular momentum for a Mestel can be analytically calculated in MOG as well as in Newtonian gravity. Therefore, in order to test our 
results, we repeated our simulations for a Mestel-like disc, and compared the results with the analytic description.
In this section for a Mestel-like disc, we do the same simulations/analysis that we did for the exponential disc. However, as we will show, the 
main results 
does not change and almost all of our analysis and interpretations will also hold for the Mestel-like disc. We recall that although the exponential
 disc is more realistic,
we also study the Mestel-like disc as a toy model, since the transfer of angular momentum for this model is possible to be interpreted analytically in MOG. Therefore, let us briefly present our results for the 
evolution of the Mestel-like disc. The initial
velocities for this disc, in both theories have been plotted in the top left panel in Fig. \ref{mestel1}. As expected, the rotation curve in MOG is higher than the Newtonian disc. It is important mentioning that this toy model does not lead to an observationally viable rotation curve.

The bar strength has been plotted in the top middle panel in Fig. \ref{mestel1}. The bar starts to form rapidly for both models
and in the Newtonian model reaches its maximum amplitude at about $\tau=1$. 
The oscillatory nature of the bar is seen again for the MOG model. For this model, the bar continues to grow with a steep slope until about $\tau=5$ and then
continues with an almost constant value. Eventually it is clear that the bar strength in MOG is weaker. More specifically, the mean time averaged value of $a_2$ in MOG is $0.21$ and in Newtonian disc is $0.29$. Therefore, the bar magnitude for our Mestel-like model in Newtonian dynamics is about $27.6 \%$ larger. 

The stability parameter $\beta$ has been illustrated in the bottom middle panel in Fig. \ref{mestel1}. This plot also verifies that the bar instability in Newtonian
disc is stronger. It is clear that $\beta$ does not get larger than $1$ at the end of simulation. It was the case also for the exponential disc in MOG. This means that for both models in MOG the random motions do not overcome the rotational motions.

In the right panel in Fig. \ref{mestel1} we have compared other Fourier modes with the bar mode for both theories. The order of the curves is the same as Fig. \ref{fig:2in1}
It is clear that, as in the case of the Mestel-like
disc, the bar mode is the most unstable mode in both  models. Also the even modes are, in general, stronger than the odd modes. It is also interesting to mention that
the $m=3$ mode is seen for both modes during the first orbital period. As we discussed before, appearance of this mode is also the case for the Mestel disc. However 
in both discs, at larger times the dominant mode is the bar mode.

For this model, the angular momentum transport by density waves in MOG has been studied in the Appendix. 

The main result of this appendix is given by equation
\eqref{a10}. This equation
is the fraction of the angular momentum that has been transported outside radius $R_0$ in a time interval $\Delta t$, to the initial angular momentum inside $R_0$.
Therefore, to compare our models, for a same time interval $\Delta t$ we can write

\begin{small}
\begin{equation}
\left(\frac{\tau_z \Delta t}{L(R_0)}\right)_{\text{M}} / \left(\frac{\tau_z \Delta t}{L(R_0)}\right)_{\text{N}}= 
\frac{\left<a_2\right>_{\text{M}}^2}{\left<a_2\right>_{\text{N}}^2}
\frac{\psi^2}{\eta}\simeq 0.65
\label{a11}
\end{equation}
\end{small}
Where subscript "M" stands for MOG and "N" for the Newtonian case. Furthermore, according to \citet{bt}, we have used the following typical values for the grand-design spirals ($m=2$): $\gamma=11^{\circ}$ and $\mu_0 R_0=0.42$. On the other hand, 
from the mean values obtained from our simulations, see Table \ref{tab1}, we have $\left<a_2\right>_{M}=0.21$  and $\left<a_2\right>_{\text{N}}=0.29$. Therefore as expected, the angular momentum transport by spiral density waves in MOG happens less effectively than the Newtonian disc. 

  For both models, $L_z$ is scaled to its initial value 
at $\tau=0$. In the Newtonian case, $L_z$ increases rapidly until about $\tau\sim 2.25$ and after that gradually continues to grow with a small slope. For the MOG model the small 
slope phase starts at a larger time, i.e. around $\tau\sim 4$. Furthermore, in this case the initial increase rate is much less, but there are more oscillations. In both cases 
$L_z$ does not stop increasing. This is expected since the stable bar rotates almost uniformly and consequently exerts an almost constant torque to the given ring.

It is also instructive to study the number of particles, i.e. $n(\tau)$, in the above mentioned region. More specifically, $n(\tau)$ is the ratio of number of the 
particles outside a circle with radius $R=0.7$ to its initial value at $\tau=0$. It is shown in the bottom left panel of Fig. 
\ref{mestel1}. The dotted black curve corresponds to the 
Newtonian disc and the dotted red curve is belonged to the MOG model. For both cases $n(\tau)$ starts to decline for a very short period of time, about $\Delta\tau\sim 0.1$. Then it 
starts to grow until about $\tau\sim 2.5$ in Newtonian disc and about $\tau\sim 4$ in the MOG model. As other physical quantities in the MOG model, $n(\tau)$ also experiences some
rapid oscillations. Since $n(\tau)$ increases with time during the bar formation, one may conclude that bar formation in our simulation is not an accretion process. In other words, 
the angular momentum of the outer disc is increased not by accreting matter into the inner disc, but by increasing the number of particles in outer parts during the bar formation 
and by increasing the circular velocity of the particles by the torque from non-symmetric density waves.

It is interesting that, in both cases, temporary behaviour of $n(\tau)$ and $L_z$ is the same. We mean when $n(\tau)$ increases/decreases, $L_z$ also increases/decreases. 
Therefore they experience a same spectrum of maximums and minimums. However $L_z$ grows forever but $n(t)$ gets approximately constant at later times.

As we have already mentioned, an important feature in the MOG model is that the bar magnitude is oscillating. This fact can also be seen for larger number of 
particles $N$. This 
means that $m=2$ pattern are somehow more transient in MOG than in Newtonian gravity. On the other hand, we know that spiral patterns can effectively
transfer the angular momentum 
throughout the disc and thus can not be long-lived patterns \citet{bt}. Now a question naturally arises: why $m=2$ patterns are more transient in MOG than in 
Newtonian disc while 
the angular momentum exchange is smaller compared to the Newtonian case? It seems that the origin of transient nature of the patterns does not depends on the angular
momentum exchange rate but rather to the way by which the density waves are excited and propagate throughout the disc.

\subsection{\small{Re-deriving Efstathiou's stability criterion}}

\begin{figure}
\centerline{\includegraphics[width=8cm]{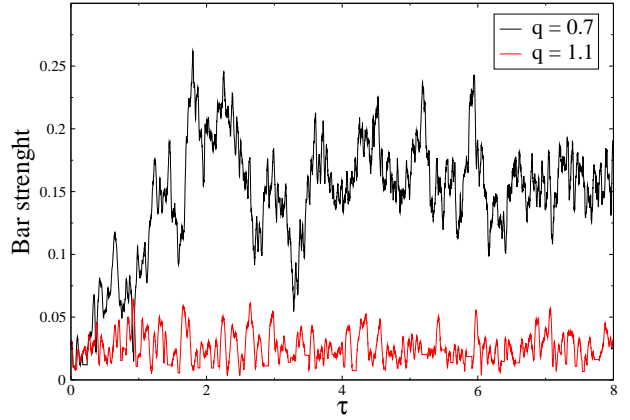}}
\caption[]{Time evolution of the bar amplitude for two different cases: $q=1.1$ and $q=0.7$.}
\label{tests}
\end{figure}

As another test, we re-drive one of the main results of \citet{ef} by adding a specific halo to our models. More specifically in \citet{ef}, it has been shown that disc is stable against bar formation if $q=V_m/(\zeta M_D G)^{1/2} \gtrsim 1.1$, where $V_m$ is the maximum rotational velocity, $M_D$ is the total disc mass and $\zeta^{-1}$ is the scale length of the exponential disc. In their main model, they have employed a rigid halo component with a density profile
\begin{eqnarray}  
\rho_H(s) &=& \zeta^3(M_D/{4\pi}) \\   \nonumber
&\times& \{ \left[ \frac{1}{2}I_1(\frac{\zeta s}{2}) K_1(\frac{\zeta s}{2})
 -\frac{3}{2} I_0(\frac{\zeta s}{2}) K_0(\frac{\zeta s}{2})\right]  \\   \nonumber
&+&\frac{\zeta s}{2} \left[ I_0(\frac{\zeta s}{2}) K_1(\frac{\zeta s}{2})-I_1(\frac{\zeta s}{2})K_0(\frac{\zeta s}{2})\right] \} \\   \nonumber
 &+&(\frac{V_m^2}{4\pi G})(r_m^2+s^2)^{-2}   \\   \nonumber
 &\times&\left[ s^2 + r_m^2(3-2\gamma)   
-\gamma(s^2+3r_m^2) \ln(\frac{s^2}{r_m^2+s^2})\right] 
\end{eqnarray}
where $\gamma=(1/2\zeta r_m)^2 /q^2$ and $r_m$ is a radius at which the rotation curve gets flat. In this model, $\zeta r_m$ and $q$ are two dimensionless parameters, which determine
the influence of the halo and consequently the behaviour of the disc against bar instability. They concluded that over the range $0.1 \leq \zeta r_m \leq 1.3$,
the disc is stable against bar formation, if $q \gtrsim 1.1$.
Adding this halo profile to a disc with exponential surface density 
\begin{equation}
 \mu_D(r)= \left( \frac{\zeta^2 M_D}{2 \pi}\right) exp(-\zeta r)
\end{equation}
results in a rotation curve 
\begin{equation}
 V_D (r) = V_m \left( \frac{r^2}{r^2 + r_m^2} \right)^{1/2} \left[ 1-\gamma \ln\left(\frac{r^2}{r^2+r_m^2} \right) \right]^{1/2}
\end{equation}
Since their calculations are performed in two dimensions (2D) , we rewrote our code in 2D and added the above mentioned halo, in order to compare the results. We tried to reproduce two of their models ($3$ and $8$), where the dimensionless parameter $\zeta r_m$ is set to $0.4$ and the $q$ is 
chosen to be $1.1$ for the stable, and $0.7$ for the unstable disc. The result has been illustrated in Fig. \ref{tests}. It is clear that the bar strength in our calculations is compatible with that of \citet{ef}. However, it should be mentioned that their method is based on a Fourier transform potential solver, which is different from ours. Therefore the behaviour of the bar strength is not exactly the same. However, the importance of the parameter $q$ as an indicator for the instability of the disc is inevitable and seen in our results. We also checked the code for different number of particles changing from 2500 to 10000, and found qualitatively same results.

\begin{figure}
\centerline{\includegraphics[width=8cm]{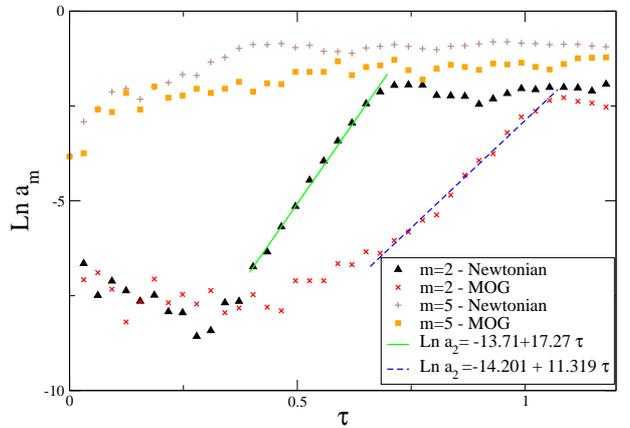}}
\caption[]{Time evolution of the amplitude of $m=5$ and $m=2$ modes on the logarithmic scale for both Newtonian and MOGian Mestel-like discs using the quiet start procedure. The solid lines shows our best fit to find the growth rates.}
\label{fig:se}
\end{figure}

\subsection{\small{Quiet start procedure}}
When the number of particles in a simulations is limited, one possible issue that can affect the results is the particle noise \citep{se89}. It is the noise in 
distribution of the particles, and results in small-scale fluctuations of density and potential. In order to confirm that our main results are independent of the 
existence of the particle noise, we checked the behaviour of the disc under a quiet start \citep{se89}. The quiet start procedure is a conventional method to suppress 
the shot noise for a while. To build the initial conditions for this method, we placed $5$ particles on each ring in our Mestel-like disc, and chose to have 1000
rings. Our main results in this section are not sensitive to the number of particles. The width of the rings are equal in order to construct a $1/R$ mass 
distribution. We randomly place the first particle of each ring. The other four particles have the same radius, and differs with each other by $2\pi /5$ in azimuthal 
angle. The amplitude of the $m=2$ mode in the logarithmic scale, for both models, has been illustrated in Fig. \ref{fig:se}. It is clear that the amplitude of this 
mode is small in both cases and start to exponentiate as $e^{\omega\tau}$ at $\tau\sim 0.37$ for the Newtonian disc and at $\tau\sim 0.6$ for the MOG model. In this 
linear regime the growth rates can be easily inferred by finding the best linear fit. The solid lines in Fig. \ref{fig:se} show our linear fits. More specifically we
find $\omega=17.27$ for the Newtonian disc and $\omega=11.32$ for the MOGian disc. It is clear that the growth rate in Newtonian model is about $0.34$\% higher than
the MOG case. This fact is compatible to our previous results. One should note that for later times, for example $\tau>0.7$ in the Newtonian case, when system enters 
the non-linear regime we can not rely on the results of the quiet start procedure.

It is necessary to mention that the above mentioned procedure for setting the initial conditions, seeds an $m=5$ mode, which grows rapidly in the disc \citep{se89}.
Therefore one should make sure that this mode does not influence the $m=2$ mode. To do so we have plotted the $m=5$ mode in Fig. \ref{fig:se}. It is clear that when
$m=5$ rapidly grows at the early times, the $m=2$ amplitude remains constant. On the other hand when $m=5$ reaches its maximum and gets constant the slope of the $m=2$
amplitude does not change. In other words, existence of the $m=5$ mode in the background does not change the exponential growth of the bar mode. This shows that the
amplitude of bar mode evolves almost independently from the $m=5$ mode.
\section{\small{Discussion \& conclusion}} 
\label{conc}

In this paper the stellar bar growth has been investigated in the context of MOG using N-body simulations for two idealised disc models. In order to make easier the 
comparison 
between MOG and Newtonian gravity, we have also performed similar simulations for a bare disc, i.e. without dark matter halo, in the Newtonian gravity. More 
specifically, we have 
followed somehow a same procedure presented in \citet{op}, where a bare Newtonian disc has been compared with a disc in which the gravitational force on the disc is
modified by assuming a rigid sphere around the disc.

Our results explicitly show that the stellar bar growth rate is significantly smaller in the MOG model. More specifically, in the case of the exponential disc, our simulations for different particles numbers reached 
to the mean value $\left<a_2\right>\sim 0.29$ in Newtonian case and $\left<a_2\right>\sim 0.17$ for the MOG model. 
This means that the final stellar bar formed in the Newtonian disc is almost $41$\% larger than the MOG case. On the other hand, for the Mestel-like disc, the bar magnitude is 
almost $27.6$ \% larger in the Newtonian Mestel-like disc. This is a 
satisfactory result for MOG considering that this theory is presented to handle the dark matter problem by ignoring the dark matter 
component of the galaxies.

Also we found out the stellar bar and spiral patterns in the MOG model are short-live structures at least in the first half of the simulation time. More specifically,
$m=2$ 
patterns are frequently excited and damped in this case. This means that excitement and propagation of the stellar density wave are, in principle, different than the
Newtonian disc. Albeit after $\tau>4$ the stellar bar settles into a stable and uniformly rotating configuration.

 Also in the case of a Mestel-like disc , we have studied the angular momentum transfer from inner parts of the disc to the outer parts in both cases. The origin for 
 this exchange is the existence of a non-symmetric
 stellar bar. We also derived an analytic expression for the angular momentum transport by non-symmetric density waves on the surface of a Mestel-like disc in MOG.
 The results show that angular momentum transfer rate in the MOG model is much less than the Newtonian case.
 
Furthermore, the pattern speed has been computed in both cases. $\Omega_p$ is slightly, is larger in MOG than in Newtonian disc.

As a matter of future study it would be instructive to compare a MOG disc with a more realistic Newtonian galaxy model by including the dark matter halo and a 
central bulge. Also 
regarding the above mentioned fact about the transient nature of the spiral patterns in MOG, it would be interesting to study the swing amplification mechanism, 
which
is one of the key features in the bar formation scenarios. Furthermore, the stability of the unbarred galaxies in MOG is of great importance in the sense that their 
stability in the Newtonian gravity has raised serious challenges.

\section*{Acknowledgements}
This work is supported by Ferdowsi University of Mashhad under Grant No. 41395(22/04/1395). Also we would like to thank Francoise Combes and Jerry Sellwood for
constructive comments, and Shahram Abbassi for providing us a high-performance computer.

\bibliographystyle{mnras}
\bibliography{paper3}

\appendix
\section{appendix}
In this appendix we explore the angular momentum exchange by an arbitrary WKB density wave in the context of MOG. Fortunately, the gravitational potential of the Mestel's disc can be derived analytically in MOG. 
Let us start with the potential and the corresponding rotation curve for Mestel disc in MOG. To do so, we mention that in the cylindrical coordinate system, solutions of \eqref{p1} and \eqref{p2} can be expended with respect to Bessel functions as, for more details see \citet{rokho} 
\begin{equation}\label{sk}
\begin{split}
& \Psi(R)=-2\pi G(1+\alpha)\int_{0}^{\infty} S(\kappa) J_0(\kappa R)d\kappa  \\ &
\chi\phi^0(R)=2\pi G\alpha\int_{0}^{\infty} \frac{S(\kappa)J_0(\kappa R)\kappa}{\sqrt{\kappa^2+\mu_0^2}} d\kappa
\end{split}
\end{equation}
where
\begin{equation}
S(\kappa)=\int_{0}^{\infty} \Sigma(R')J_0(\kappa R')R' dR'
\end{equation}
the surface density of a Mestel's disc is $\Sigma=\sigma_0 r_0/R$, where $r_0$ is a scale length. In this case it $S(\kappa)=\sigma_0 r_0/\kappa$. Integrals in equation \eqref{sk} can be simply solved and the effective potential $\Phi$ for the Mestel's disc in MOG takes the following form
\begin{equation}\label{sk1}
\Phi(R)=v_0^2(1+\alpha) \ln R+v_0^2\alpha I_0(\frac{\mu_0 R}{2})K_0(\frac{\mu_0 R}{2})
\end{equation}
In this equation $I_n$ and $K_n$ are modified Bessel functions of the first and second kinds. Setting $\alpha$ to zero, the Newtonian potential for this disc is recovered. Now one may straightforwardly show that the rotation curve, $v^2=R d\Phi/dR$, is
\begin{equation}
v(x)=v_0\sqrt{(1+\alpha)+\alpha x \left( I_1(x)K_0(x)-I_0(x) K_1(x)\right)}
\label{a7}
\end{equation}
where $x=\mu_0 R/2$. As expected when $\alpha=0$ the rotation curve is constant everywhere.

Now following a same procedure presented in \citet{bt}, we drive a relation for the angular momentum transport by the density waves, like spiral arms, in MOG. The $z$ component of the torque exerted on the material outside a cylinder with radius $R_0$ is written as 
\begin{equation}
\tau_z=\int_{R_0}^{\infty}\int_0^{2\pi}\int_{-\infty}^{\infty}  \frac{\partial\Phi}{\partial\varphi}\rho R dR d\varphi dz
\label{a1}
\end{equation}
where $\rho$ is the matter density and can be substituted from the modified Poisson equation (\ref{n1}). Using some iterations and applying the divergence theorem, we find
\begin{equation}
\rho\simeq\frac{\nabla^2 \Phi}{4\pi G} -\frac{\alpha \mu_0^2}{16 \pi^2 G} \int \Phi(r') \nabla^2\big[\frac{e^{-\mu_0 |r -r'|}}{|r-r'|}\big]d^3x'
\label{a2}
\end{equation}
where we kept only the terms linear in $\mu_0^2$. One may still simplify equation \eqref{a2} and after some manipulations derive the final form  
\begin{equation}
\rho\simeq\frac{\nabla^2 \Phi}{4\pi G} -\frac{\alpha \mu_0^2}{16 \pi^2 G} \Phi
\label{a3}
\end{equation}
Substituting this equation into \eqref{a1} and using the fact that $\int_0^{2\pi} \Phi \frac{\partial \Phi}{\partial \varphi} d\varphi=0$, we find
\begin{equation}
\tau_z=\frac{R_0}{4\pi G}\int_{-\infty}^{\infty}\int_0^{2 \pi} R_0 \left(\frac{\partial \Phi}{\partial \varphi} \frac{\partial \Phi}{\partial R}\right)_{R=R_0} dz d\varphi
\label{aa1}
\end{equation}
In fact the second term in \eqref{a3} does not contribute in $\tau_z$. In other words, equation \eqref{aa1} is the same equation as in the Newtonian gravity. However it should be noted that $\Phi$ is different from the Newtonian case and consequently the angular momentum exchange rate would be different. 

Now let us assume a WKB density wave $\Sigma_1(R, \phi,t)$ on the surface of the disc given by
\begin{equation}
\Sigma_1=H(R)\cos(m\phi+f(R,t))
 \end{equation}
where $H$ is a slowly varying function of radius and $f(R,t)$ is the shape function. This function is related to the radial wave number as $k=\partial f/\partial R$. 
This wave have a $m$-fold rotational symmetry and $m>0$ arms. The dispersion relation of this perturbation in the context of MOG and its gravitational potential have 
been investigated in \citet{ro2015}. The corresponding gravitational potential is
\begin{equation}
\Phi_1=-\frac{2\pi G}{|k|}\Sigma_1 \psi
\end{equation}
where
\begin{equation}
\psi=1+\alpha-\frac{|k|\alpha}{\sqrt{k^2+\mu_0^2}} >1
\label{a5}
\end{equation}
regarding the linearity of the modified Poisson equations, the total potential of the disc and the perturbation can be written as $\Phi=\Phi_0(R)+\Phi_1$, where $\Phi_0(R)$ is the potential of the background disc and is given by equation \eqref{sk1}. Finally, using \eqref{aa1} $\tau_z$ can be written as
\begin{equation}
\tau_z=\text{sgn}(k) \frac{\pi^2 m R_0 G a_m^2 \Sigma(R_0)^2 }{k^2} \psi^2  
\label{a4}
\end{equation}
where we have assumed $H=a_m \Sigma$, and $a_m$ is the Fourier amplitude. The radial wave number can be written with respect to the pitch angle $\gamma$ as $k=\frac{m}{R\tan\gamma}$. To investigate the strength of this torque, let us calculate the total angular momentum inside the radius $R=R_0$ in the unperturbed disc, i.e.
\begin{equation}
L(R_0)= 2\pi \int_0^{R_0} R^2 v(R) \Sigma(R) dR
\label{a6}
\end{equation}
For the Mestel's disc, by assuming that $\mu_0 R_0 \ll 1$ and using the rotation curve \eqref{a7}, we find
\begin{equation}
L(R_0)=\frac{v_0^3 R_0^2}{2G}\eta(\alpha,x_0)
\label{a8}
\end{equation}
where $x_0=\mu_0 R_0/2$ and
\begin{equation}
\eta(\alpha,x_0)=1+\frac{\alpha x_0^2}{4}\left( 1.06-\ln 2x_0\right)>1 
\label{a9}
\end{equation}
On the other hand, the angular momentum transferred to the outer disc in a time interval $\Delta t$ is $\Delta L_z\sim \tau _z\Delta t$. In this case, the fraction of
the angular momentum that has been transported outside radius $R_0$ in a time interval $\Delta t$ to the initial angular momentum inside radius $R_0$ is
\begin{equation}
\frac{\tau_z \Delta t}{L(R_0)}=\text{sgn}(k)\frac{\pi \tan^2\gamma N_{\text{rot}} a_m^2}{m}\left(\frac{\psi}{\eta}\right)^2
 \label{a10}
\end{equation}
where $N_{\text{rot}}= \frac{v(R_0) \Delta t}{2\pi R_0}$ is the total number of the rotations in this time interval. By setting $\psi$ and $\eta$ to unity, the Newtonian expression is recovered.

\bsp
\label{lastpage}
\end{document}